\documentclass[aps,prb,superscriptaddress,preprintnumbers,
showpacs,A4paper,twoside,twocolumn]{revtex4-1}

\usepackage{bm}
\usepackage{graphicx}
\usepackage{subfigure}
\usepackage{amsfonts}
\usepackage{amsmath}
\usepackage{amssymb}
\begin{document}

\title{Cluster statistics of critical Ising and Ashkin-Teller models}

\author{Zhaonian Xu}
\affiliation{Department of Physics and Beijing Key Laboratory of Opto-electronic Functional Materials \& Micro-nano Devices, Renmin University of China, Beijing 100872, China}

\author{Rong Yu}
\affiliation{Department of Physics and Beijing Key Laboratory of Opto-electronic Functional Materials \& Micro-nano Devices, Renmin University of China, Beijing 100872, China}

\affiliation{Key Laboratory of Quantum State Construction and Manipulation (Ministry of Education), Renmin University of China, Beijing, 100872, China}


\begin{abstract}
Motivated by recent progress on the scaling behavior of entanglement entropy, we study the scaling behavior of the number of clusters crossing the boundary between two subsystems for several classical statistical models in two dimension. This number exhibits a subleading logarithmic dependence of the linear dimension of the boundary when the model is at critical, in analogy to the entanglement entropy of a quantum system. It is shown that the logarithmic scaling of the cluster number originates from the conformal invariance of the critical system. We check this numerically for Ising and Ashkin-Teller models by using Monte Carlo simulations, and show that whether a universal coefficient of the logarithmic term can be observed numerically may strongly depend on the geometry and boundary condictions of the system. 
\end{abstract}

\maketitle

\section{Introduction}
Entanglement is a fundamental concept in quantum physics. As a measure of quantum information inherent in a quantum state, the entanglement entropy has been extensively studied in many areas of physics~\cite{ref1}. In condensed matter physics, it has been shown that
the scaling behavior of the entanglement entropy is able to characterize the nature of a quantum phase transition (QPT)~\cite{ref2,ref3}. 
Assuming a system consisting of two subsystems A and B. The von Neumann entanglement entropy is defined
by
\begin{equation}\label{Eq:EE_Def}
S=-\rm{Tr}\rho _A \log_2\rho _A,
\end{equation}
where $\rho _A$ is the reduced density matrix of subsystem A, which is obtained from the total density matrix by tracing out degrees of freedom of subsystem B. Note that the R\'{e}nyi entropy, which is another quantity frequently used in disussing the entanglement, can be defined in a similar way. Denoting the linear dimensions of the subsystem A and the combined system $\rm{A}\cup \rm{B}$ as $L$ and $M$, respectively, many studies have shown that when $L\ll M$, the entanglement entropy of the ground state of a quantum many-body system takes the following form\cite{ref4}
\begin{equation}\label{Eq:EE_Scaling}
 S(L) = aL^{d-1} + b\ln L + k,
\end{equation}
where $d$ refers to the dimensionality of the subsystem A.
In general, the leading term of $S(L)$ follows an area law (proportional to the area of the boundary of subsystem A). When the system is critical, an additional logarithmic correction appears as the subleading term. In Eq.~\eqref{Eq:EE_Scaling}, the coefficient $a$ is proportional to the boundary free energy of the subsystem A which depends on microscopic detail of the system. The constant $k$ is shown to be universal for systems possessing a topological order~\cite{ref5,ref6}. For $(1+1)D$ critical systems, the coefficient $b$ of the logarithmic term is proportional to the central charge $c$ of the underlying conformal field theory (CFT)~\cite{ref7,ref8,ref9}. Similar conclusion holds for a class of conformal quantum critical models in higher dimension, but the geometry of the subsystem A is relevant and can give rise to corner correction to the coefficient $b$.\cite{ref10} Nonetheless, it is generally believed $b$ is universally determined by the underlying CFT that controls the phase transition\cite{ref11}.   

Recent numerical results of the entanglement entropy at the QPT between a plaquette valence bond solid (PVBS) and an antiferromagnetic (AFM) phase in a $J$-$Q$ model questions the above general belief\cite{ref12}. The transition is in proximity to a deconfined quantum critical point (DQCP) with emergent $SO(5)$ symmetry~\cite{Sandvik_2007}. Though the putative DQCP can be described by the $SO(5)$ CFT, numerical evidences suggest that the coefficint $b$ at the PVBS-to-AFM transition strongly depends on the geometry of the subsystem~\cite{ref12,Meng_2023,Sandvik_2024}.  

Though interesting, a full answer to the above question is difficult given the complication of the quantum system. Here we study a similar problem in several $2D$ classical statistical models. Still assuming the system can be divided into A and B subsystems. We simply count the number of spin clusters $N(L)$ that touch the bounary between the two subsystems. This number is closely related to the entanglement entropy. It is shown that in random transverse field Ising models $S(L)$ is proportional to $N(L)$\cite{ref13}. This is because the partition function of the system can be written as a summation over independent clusters, and each cluster is in a GHZ state which contributes equally ($\log 2$) to $S(L)$. 

In this paper we show that 
\begin{equation}\label{Eq:Nc_Scaling}
N(L) = aL + b\ln L
\end{equation}
holds for a class of classical spin models at critical. For $Q$-state Potts models, the analytical results indicate that the coefficient $b$ can be determined by underlying boundary CFT. This is further checked by our Monte Carlo (MC) simulations at the critical temperature of the Ising model. Interestingly, our results on the Ashkin-Teller model (ATM) suggest a non-universal behavior of $b$ varying with the model parameter. We further show that both the subsystem geometry and boundary condictions can strongly affect the value of the coefficient $b$. We hope our results shed some light in understanding the logarithmic behavior of entanglement entropy in quantum critical systems.

\section{Models and some analytical results}
In this paper we consider two statistical models defined on a $2D$ square lattice.
The first one is the Ising model, whose Hamiltonian reads as
\begin{equation}\label{Eq:Ising}
 H = -\frac{J}{2} \sum_{i,\delta} \sigma_i \sigma_{i+\delta},
\end{equation}
where $\sigma_i=\pm1$ is the Ising variable on the $i$-th lattice site, $\delta$ runs over all nearest neighbors, and $J>0$ is the ferromagnetic coupling. We consider the model at the critical temperature $T_c$ of the magnetic transition, where the system is described by a $c=1/2$ CFT.
The other model we consider is the ATM defined on a bilayer lattice: 
\begin{equation}\label{Eq:ATM}
H = -\frac{J}{2} \sum_{i,\delta} (\sigma_i \sigma_{i+\delta} + \tau_i \tau_{i+\delta}) - K  \sum_{i,\delta} \eta_i \eta_{i+\delta},
\end{equation}
where $\sigma_i$ and $\tau_i$ are Ising variables defined on sites of the top and bottom layers, respectively. $\eta_i=\sigma_i \tau_i$ is also an Ising variable. For $0\leqslant K/J \leqslant 1$, the model has a continuous phase transition described by a $c=1$ CFT on a $Z_2$ orbifold. For $K=0$, the model is identical to two decoupled Ising models, and for $K=J$, the transition is in the $4$-state Potts universality. However, for $0<K/J<1$ the transition exhibits non-universal critical exponents varying with the model parameter $K/J$. 
For convenience, in the rest of the paper, we have set $J=1$ in Eqs.~\eqref{Eq:Ising} and ~\eqref{Eq:ATM}. 

\begin{figure}[h]
	\begin{center}
		\includegraphics[
		width=85mm,angle=0]{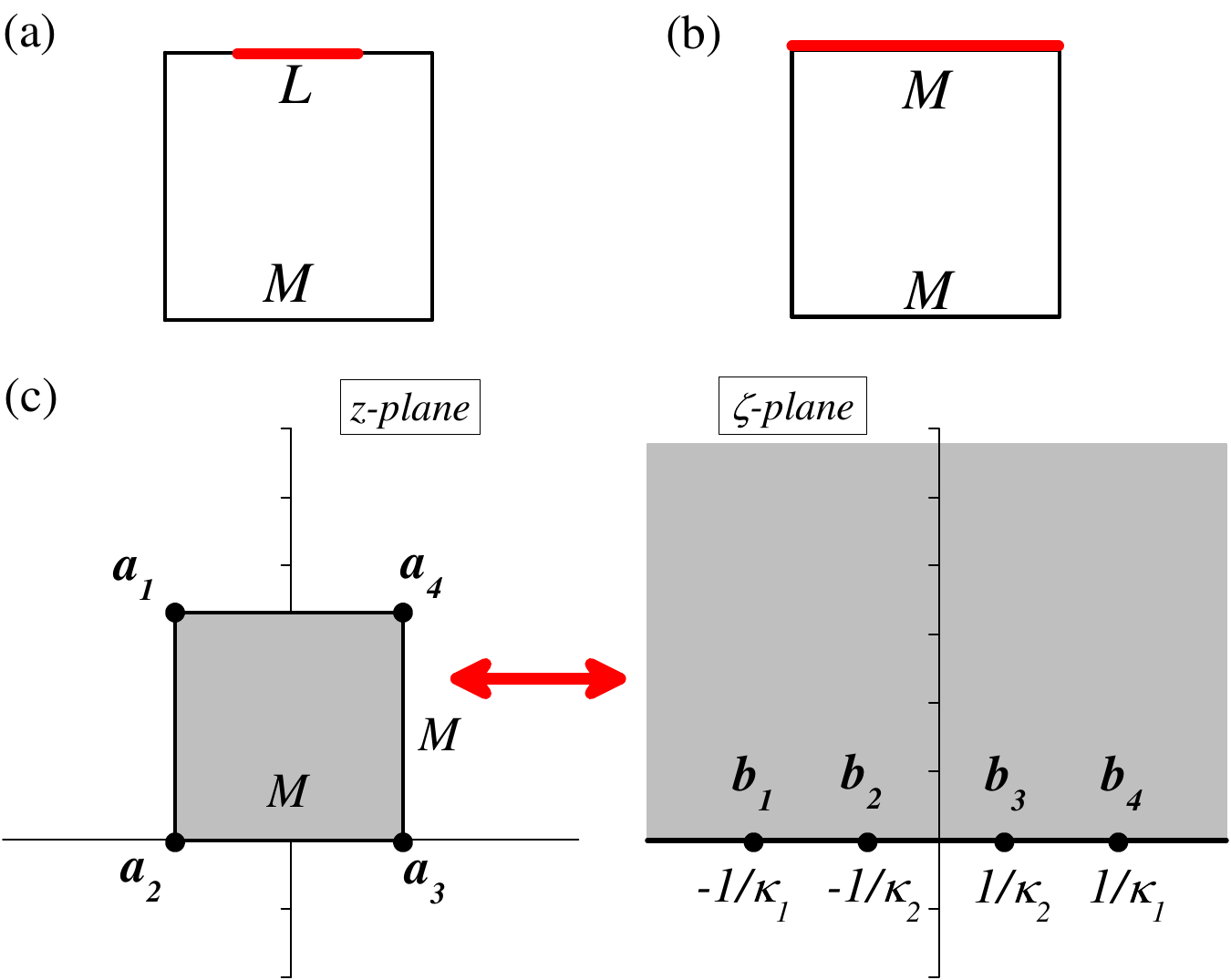}
		\caption{(a) A sketch of the system (a box with linear size $M$) and subsystem A, which is a line segment with length $L$ on the boundary of the box. (b) Similar to (a) but the subsystem A is one edge of the box with length $M$. (c) Illustration of the conformal mapping between the box with linear size $M$ and the uppper half plane.} \label{fig:system}
	\end{center}
\end{figure}

To obtain the scaling behavior of the cluster number $N(L)$, the basic idea is to conformally map a
complex geometry to the upper half plane, with its boundary to be transformed to the real axis. We can then define primary fields related to the number of clusters touching the boundary. In the following, we consider two cases where the subsystem A are defined as line segments as illustrated in Fig.~\ref{fig:system}(a) and (b). 

The conformal mapping connecting the finite square to the half plane
is, according to the Schwarz-Christoffel transformation,
\begin{widetext}
\begin{eqnarray}\label{E.ConfMapSquare}
z = \int d\zeta (\zeta+\frac{1}{\kappa_1})^{-\frac{1}{2}}
(\zeta+\frac{1}{\kappa_2})^{-\frac{1}{2}}
(\zeta-\frac{1}{\kappa_2})^{-\frac{1}{2}}
(\zeta-\frac{1}{\kappa_1})^{-\frac{1}{2}}.
\end{eqnarray}
\end{widetext}
It maps the upper half plane to the interior of a square, with the
real axis to the four edges. Especially we find the point $b_i$ on
the real axis of the $\zeta$-plane is mapped to a vertex $a_i$ of the square
in the $z$-plane (see Fig.~\ref{fig:system}(c) for coordinates of these
points). Parameters $\kappa_1$ and $\kappa_2$ can be determined by
plugging in the coordinates of $a_3$ and $a_4$:
\begin{eqnarray}
\frac{M}{2} = \kappa_1\kappa_2 \int^{\frac{1}{\kappa_2}}_0
\frac{d\zeta}{\sqrt{(1-\kappa^2_1\zeta^2)(1-\kappa^2_2\zeta^2)}}\\
\frac{M}{2}+iM = \kappa_1\kappa_2 \int^{\frac{1}{\kappa_1}}_0
\frac{d\zeta}{\sqrt{(1-\kappa^2_1\zeta^2)(1-\kappa^2_2\zeta^2)}},
\end{eqnarray}
which leads to
\begin{eqnarray}
\kappa_1 K(\frac{\kappa_1}{\kappa_2}) = \frac{M}{2}\\
\kappa_1 K(\sqrt{1-\frac{\kappa_1}{\kappa_2}}) = M,
\end{eqnarray}
where $K(x)=\int_0^1 \frac{d\zeta}
{\sqrt{(1-\zeta^2)(1-x^2\zeta^2)}}$, is the elliptic integral of
the first kind. We then immediately obtain
\begin{equation}
\kappa_1 \propto \kappa_2 \propto M.
\end{equation}

We then consider a general $Q$-state Potts model\cite{ref14} defined on the upper half plane of $\zeta$, and the Ising model corresponds to $Q=2$. It is well known that the partition function of the critical $Q$-state Potts model can be expressed as a sum over clusters, or
equivalently, dense loops~\cite{ref15, Nienhuis_book}
\begin{equation}
Z = \sum_P \sqrt{Q}^{N_P}.
\end{equation}
At the critical point, each loop gets the same weight
$\sqrt{Q}$. We now introduce a boundary conditions changing operator
$\phi_y$, whose two point function of is defined by modifying the weights from $\sqrt{Q}$ to $y$ for loops touching the boundary between the two insertions $\zeta_1$ and $\zeta_2$:
\begin{equation}
\langle \phi_y(\zeta_1)\phi_y(\zeta_2) \rangle = \frac{1}{Z} \sum
\sqrt{Q}^{N_P} \left(\frac{y}{\sqrt{Q}}\right)^{N^b_P(\zeta_1,\zeta_2)},
\end{equation}
where $N^b_P(\zeta_1,\zeta_2)$ gives the number of loops touching
the boundary between $\zeta_1$ and $\zeta_2$ located on the real
axis. Given the conformal invariance, we expect the two point function scales as
\begin{equation}\label{E.scaling}
\langle \phi_y(\zeta_1)\phi_y(\zeta_2) \rangle \sim e^{-f(y)|\zeta_1-\zeta_2|}
|\zeta_1-\zeta_2|^{-2h(y)},
\end{equation}
where $f(y)$ is the boundary free energy induced by the modified
weight on the boundary, and the exponent $h(y)$ is the anomalous
dimension of the operator $\phi$.
Now differentiate the two point function of boundary conditions
changing operators with respect to the weight $y$, then take the
limit $y=\sqrt{Q}$. This leads to
\begin{equation}
N^b_P(l) = al + b\ln l,
\end{equation}
where
\begin{eqnarray}
a &=& -\sqrt{Q} \left[\frac{\partial f}{\partial y}\right]_{y=\sqrt{Q}}, \\
b &=& -2\sqrt{Q}\left[\frac{\partial h(y)}{\partial
	y}\right]_{y=\sqrt{Q}},\label{E.bdiff}
\end{eqnarray}
and $l=|\zeta_1-\zeta_2|$.
With the exact form of $h(y)$,\cite{JS}
we obtain
\begin{equation}
b = \frac{1}{2\pi p} \sqrt{Q(4-Q)} \label{E.prefactor}
\end{equation}
where we parameterize
\begin{equation}
\sqrt{Q} = 2\cos \frac{\pi}{p+1},
\end{equation}
and the central charge 
\begin{equation}
 c = 1-\frac{6}{p(p+1)}.
\end{equation}

We then go back to the problem of the sqaure box defined in the $z$-plane. 
For a primary field
$\phi$ with dimension $h$,
\begin{equation}\label{E.PrFieldTransf}
\phi(\zeta_1,\zeta_2) = \left(\frac{\partial z_1}{\partial
	\zeta_1}\right)^h \left(\frac{\partial z_2}{\partial
	\zeta_2}\right)^h \phi(z_1,z_2).
\end{equation}
From Eq.~\ref{E.ConfMapSquare}, the derivative of $z$ with respect
to $\zeta$ is
\begin{equation}\label{E.Jacobian}
\frac{\partial z}{\partial \zeta} = \frac{\kappa_1\kappa_2}
{\sqrt{(1-\kappa^2_1\zeta^2)(1-\kappa^2_2\zeta^2)}}.
\end{equation}
Note that $\phi(\zeta_1,\zeta_2)\sim |\zeta_1-\zeta_2|^{-2h}$, 
we then have
\begin{widetext}
\begin{equation}\label{E.FieldScaling}
\phi(z_1,z_2)\sim
\kappa_1^{-2h}\kappa_2^{-2h}(1-\kappa^2_1\zeta^2_1)^{\frac{h}{2}}
(1-\kappa^2_2\zeta^2_1)^{\frac{h}{2}}
(1-\kappa^2_1\zeta^2_2)^{\frac{h}{2}}
(1-\kappa^2_2\zeta^2_2)^{\frac{h}{2}} |\zeta_1-\zeta_2|^{-2h}.
\end{equation}
\end{widetext}

We first consider the case where the subsystem A is a line segment with length $L=|z_1-z_2|\ll M$, as illustrated in Fig.~\ref{fig:system}(a). We expand $z_1$ around $z_2$ where $[\partial z/\partial
\zeta]_{z_2}\neq 0$ or $\infty$:
\begin{equation}\label{E.LinearExpansion}
z_1-z_2 = \left[\frac{\partial z_1}{\partial
	\zeta_1}\right]_{\zeta_2} (\zeta_1-\zeta_2).
\end{equation}
Then from Eq.~\eqref{E.PrFieldTransf}, we find
\begin{equation}
\phi(z_1,z_2) \sim |z_1-z_2|^{-2h}.
\end{equation}
Therefore, the anomalous dimension $h(y)$ is not affected by the conformal transformation, and we find 
the cluster number follows Eq.~\eqref{Eq:Nc_Scaling} where the coefficient $b$ is determined to be as in Eq.~\eqref{E.prefactor}.

We then consider the other case where the subsystem A takes one edge of the box with length $M$.
Note that $L$ is now fully determined by $M$.
So we expect $\phi(z_1,z_2)\sim M^{-2h^\prime}$, which leads to
a term proportional to $\ln M$ for the cluster number $N(M)$. Expanding
$z_1$ and $z_2$ at the vertexes $a_3$ and $a_2$, respectively, we get
\begin{eqnarray}
1-\kappa_2\zeta_1 =
2(1-\frac{\kappa^2_1}{\kappa^2_2})(\frac{z_1-M/2}{\kappa_1})^2;\\
1+\kappa_2\zeta_2 =
2(1-\frac{\kappa^2_1}{\kappa^2_2})(\frac{z_2+M/2}{\kappa_1})^2.
\end{eqnarray}
Then from Eq.~\eqref{E.FieldScaling} we obtain
\begin{eqnarray}
\phi(z_1,z_2) &\sim& \kappa_1^{-2h}\kappa_2^{-2h}
(1-\kappa_2\zeta_1)^{h/2} (1+\kappa_2\zeta_2)^{h/2}
(2/\kappa_2)^{-2h}\nonumber\\
&\sim& \kappa_1^{-2h} (\frac{z_1-M/2}{\kappa_1})^h
(\frac{z_2+M/2}{\kappa_1})^h\nonumber\\
&\sim& M^{-4h}.
\end{eqnarray}
Therefore, $h^\prime = 2h$. As a consequence,
\begin{equation}
 N(M) = a^\prime M +b^\prime \ln M,
\end{equation}
where $b^\prime=2b$ and $b$ is the value in Eq.~\eqref{E.prefactor}.

\section{Numerical Results}
In this section, we study the scaling behavior of $N(L)$ by studying the critical Ising model and ATM by using MC simulations with cluster algorithms~\cite{SwedsenWang, ref16}. Though there are subtle differences between these two algorithms, they share the same way in generating clusters, 
and the clusters percolate at the Ising transition temperature $T=T_c$. 
Taking $Q=2$ for the Ising model, we obtain from Eq.~\eqref{E.prefactor} 
$b = 1/(3\pi)\approx 0.106$.
On the other hand, for the 4-state Potts model, taking $Q=4$ in Eq.~\eqref{E.prefactor} we get $b=0$, \textit{i.e.}, no logarithmic correction.

We can check these analytical results via MC simulations. The clusters are constructed during MC updates, and at each MC step we count the cluster number $N(L)$ touching the boundary of a given subsystem with linear size $L$. In all simulations we find a non-universal leading term of $N(L)\sim aL$. To better resolve the subleading logrithmic correction, we define 
\begin{equation}
\delta N (L) = 2N (L) - N(2L).
\end{equation} 
With the scaling form in Eq.~\eqref{Eq:Nc_Scaling} for $N(L)$, we expect
\begin{equation}
\delta N (L) \sim \ln L + \rm{const}.
\end{equation} 

\begin{figure}[h]
	\centering
	\includegraphics[
	width=85mm,angle=0]{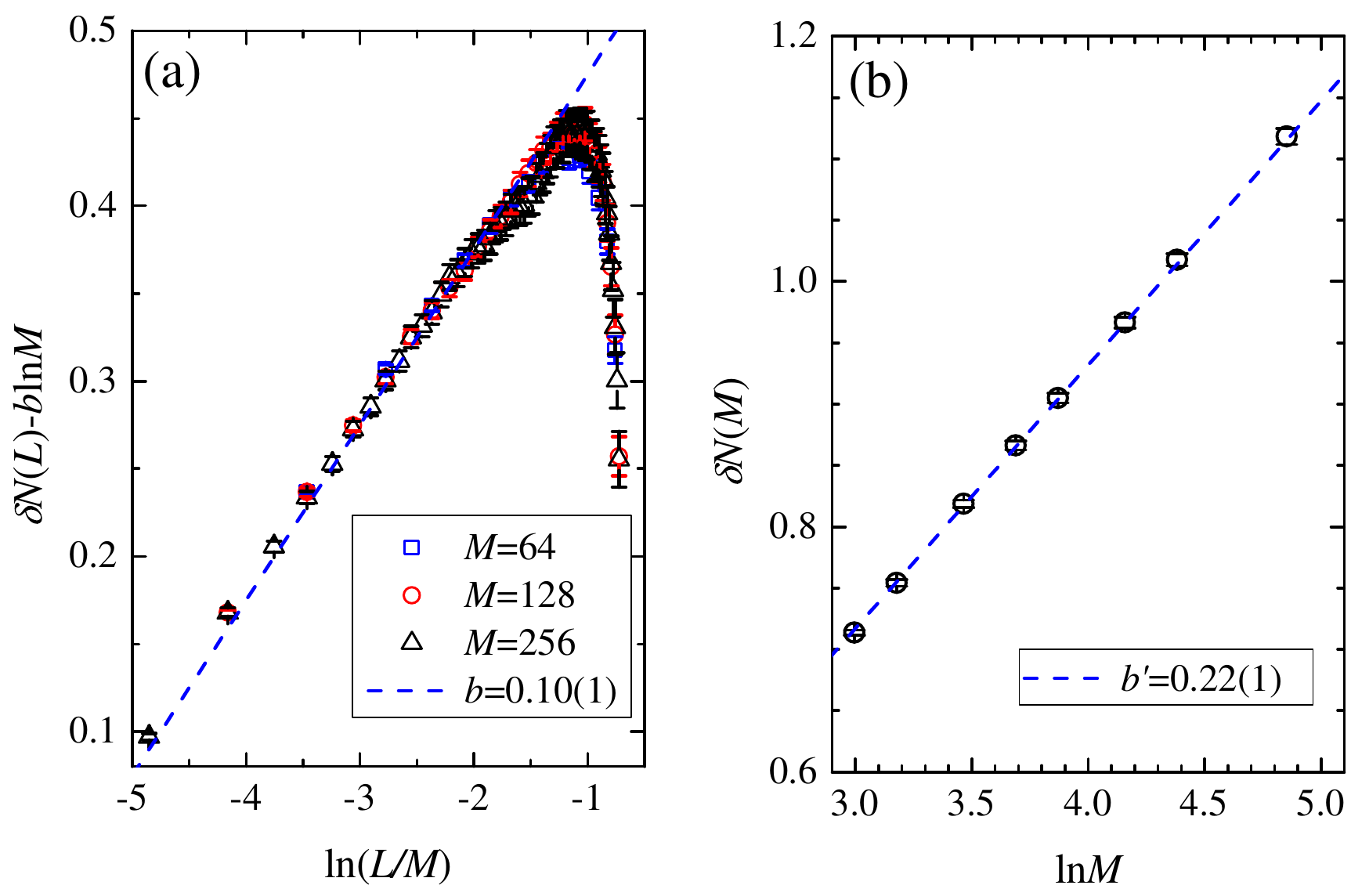}
	\caption{Finite-size scaling of $\delta N$ in the critical Ising model. The subsystems in (a) and (b) are illustrated in Fig.~\ref{fig:system}(a) and (b), respectively. In each case, the blue dashed line is a best fit of the logarithmic term, and the extracted coefficient consistent with the analytical value.} \label{fig:IsingLine}
\end{figure}

In Fig.~\ref{fig:IsingLine} we show numerical results of the scaling of $\delta N$ in the critical Ising model. The subsystems used for results shown in panels (a) and (b) correspond to those shown in Fig.~\ref{fig:system}(a) and (b), respectively. In both cases, the extracted $b$ and $b^\prime$ values are consistent with the analytical ones. Especially, the $b^\prime$ value in case (b) is about twice of that in case (a), which is a natural consequence of conformal invariance. However, we find that $b$ deviates from the value $c/3$ which is for the entanglement entropy in the $(1+1)D$ case. Although in our case the central charge of the underlying CFT should also be relevant, the $b$ coefficient is more obviously connected to the anomalous dimsnion of the boundary conditions changing operator $\phi$, as shown in Eq.~\eqref{E.bdiff}. Moreover, the doubling of $b$ value in case (b) indicates that depending on the geometry of the subsystem A, the Jacobian in the conformal transformation (see Eq.~\eqref{E.ConfMapSquare}) may also contribute to the anomalous dimension of the associated operator and hence modify the $b$ coefficient. As a further remark, we see from Fig.~\ref{fig:IsingLine}(a) that the correct $b$ value is only observed when $L\ll M$, although finite-size scaling behavior is still observed for sizable $L/M \gtrsim 1/3$. 
  
\begin{figure}[h]
	   \centering
	   \includegraphics[
	   width=85mm,angle=0]{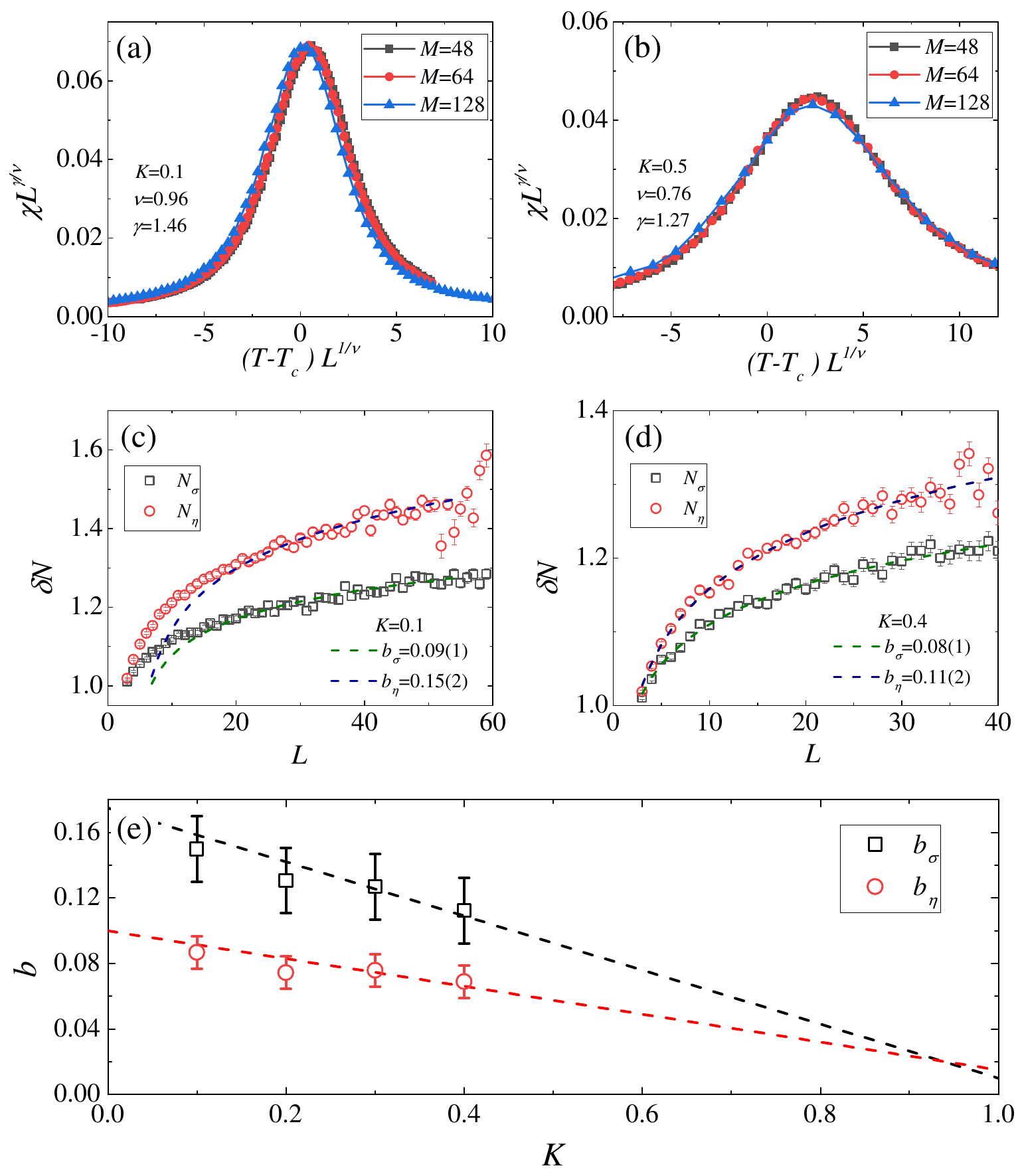}
	\caption{(a), (b) Finite-size scaling of the susceptibility of the critical ATM at $K=0.1$ and $K=0.5$, respectively. The extracted critical exponents $\nu$ and $\gamma$ vary with $K$ and are non-universal. (c) Scaling of $\delta N(L)$ for clusters associated with $\sigma$ and $\eta$ in the ATM at $K=0.1$. Dashed lines are logarithmic fits. (d) Same as (c) but at $K=0.4$. (e) Evolution of the coefficient $b$ with $K$. Dashed lines are linear fits.
	 }
	\label{fig:ATM}
	\end{figure}

\begin{figure}[h]
	\centering
	\includegraphics[
	width=85mm,angle=0]{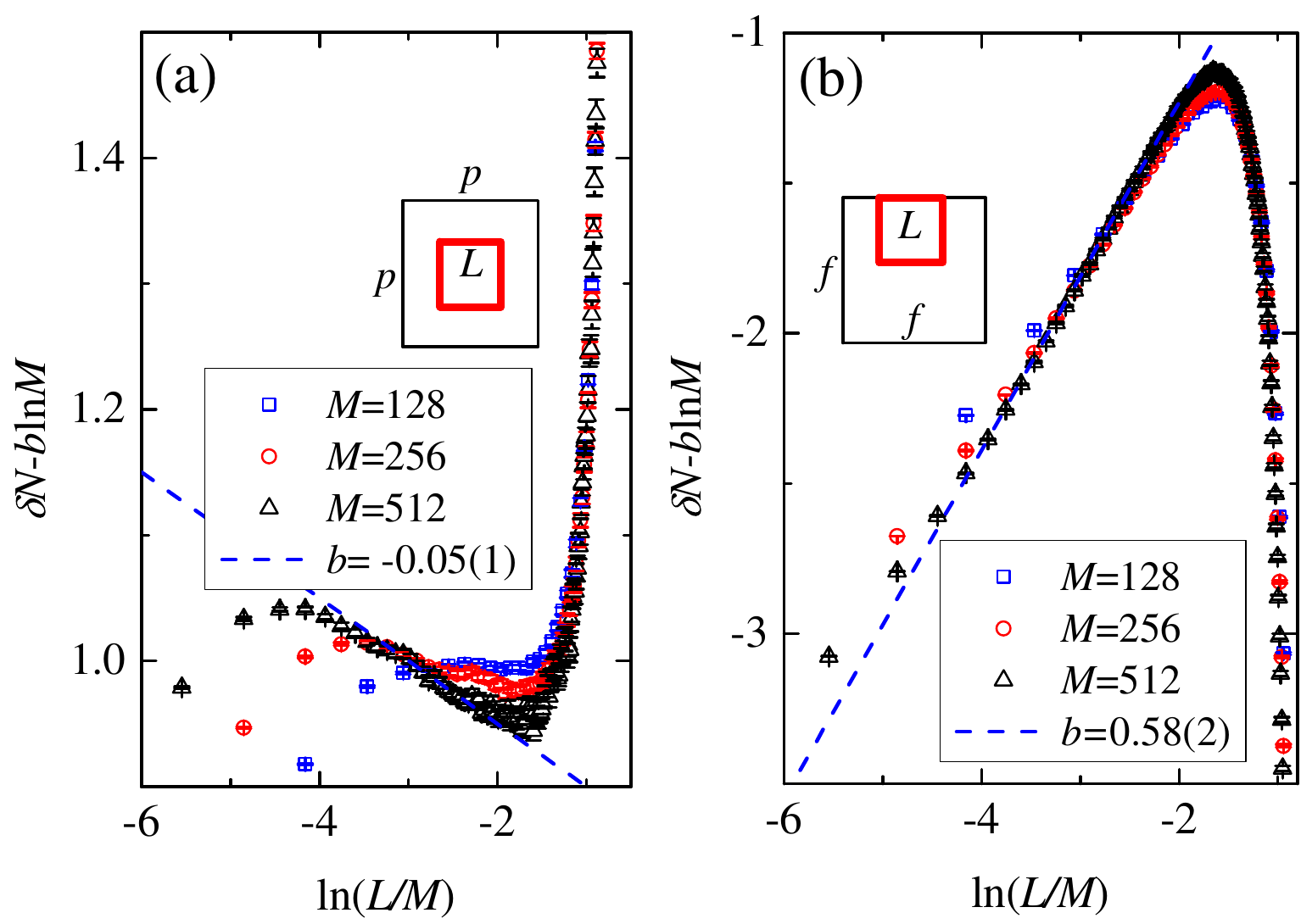}
	\caption{Finite-size scaling of $\delta N$ in the critical Ising model. In each case, the subsystem is a box with length $L$. Periodic [in (a)] and free (open) [in (b)] boundary conditions of the total system are applied in (a) and (b), respectively. In each case, the blue dashed line is a best fit of the logarithmic term with coefficient $b$.
	}
	\label{fig:IsingBox}
\end{figure}

We then study the scaling of $N(L)$ in the ATM. It is well known that the criticality is controlled by an underlying CFT with $c=1$ on a $Z_2$ orbifold. The critical exponents vary with the coupling $K$ and are non-universal, as shown in Fig.~\ref{fig:ATM}(a) and (b). Interestingly, we find the coefficient $b$ of the logarithmic term in $N(L)$ is also $K$ dependent as shown in Fig.~\ref{fig:ATM}(c) and (d). The fitted $b$ value decreases with increasing $K$.  
Besides the $K$ dependence, we find $\delta N(L)$ associated with spin variables $\sigma$ and $\eta$ also scale differently. For $K=0.1$, we find $b_{\sigma}\approx 0.09$, $b_{\eta}\approx 0.15$, and roughly $b_{\eta}/b_{\sigma}\sim 2$. But for $K=0.4$, $b_{\sigma}\approx 0.08$, $b_{\eta}\approx 0.11$, and clearly $b_{\eta}/b_{\sigma}<2$.
These can be understood as follows. In the Ising limit $K\rightarrow0$, given that $\eta=\sigma\tau$, the scaling dimension associated with the operator for $\eta$ should be twice as that for $\sigma$. Therefore, in this limit, we expect $b_{\sigma}=1/(3\pi)$ and $b_{\eta}/b_{\sigma}=2$. On the other hand, in the $4$-state Potts limit $K\rightarrow1$, the system has an enhanced $S_4$ permutation symmetry. As a consequence of this symmetry, $\sigma$, $\tau$, and $\eta$ should have the same scaling dimension. Moreover, by taking $Q=4$ in Eq.~\eqref{E.prefactor}, we expect $b_{\eta}=b_{\sigma}=0$ in this limit. With these analysis, the numerical results at $K=0.1$ can be understood as proximity to the Ising limit, and the decrease of both $b_\sigma$ and $b_\eta$ with increasing $K$ reflects the evolution of the system from the Ising universality to the Potts one. Once again, we see here the more relevant factor to the coefficient $b$ is not the central charge but the scaling dimension of the corresponding operator.    

Before we conclude, we go back to the critical Ising model and consider the subsystem A to be a box with length $L$. In Fig.~\ref{fig:IsingBox} we show the scaling of $\delta N(L)$ for two cases where the total system (a box with length $M>L$) takes periodic [in panel (a)] and free (open) [in panel (b)] condition. Surprisingly, we find the coefficient $b$ in these two cases takes very different values, and even with opposite signs. Although for this subsystem geometry we cannot obtain analytical results, we point out that the positive $b$ value in panel (b) of Fig.~\ref{fig:IsingBox}, where the system has an open boundary condition, implies that the boundary conditions changing operator defined in Section II is highly relevant, in addition to any corner correction.  

\section{Conclusion}

In conclusion, we have analyzed the cluster number crosses the boundary of a subsystem in Ising and Ashkin-Teller models at critical. By combining analytical and numerical techniques, we find that this number exhibits a subleading logrithmic term with the subsystem length $L$, and the coefficient $b$ of this term is directly related to the anomalous dimsnsion of the corresponding operator instead of the central charge of the underlying conformal field theory. We also find that besides the corner correction, the boundary condition is also a highly relevant factor to this logarithmic term. Given the analogy between the cluster number and the entanglement entropy, we hope our results help solving some puzzles in the scaling behavior of the entanglement entropy in some quantum critical systems.  

\textit{Acknowledgments.}
We thank H. Saleur for useful discussions. This work is supported by the National Key R\&D Program of China (Grant No. 2023YFA1406500), and the National Science Foundation of China (Grant Nos. 12334008 and 12174441).

\end{document}